\documentclass{article}

\usepackage{graphics}
\usepackage{amssymb, amsmath}

\newcommand{\LI}
	{\begin{list}{$\bullet$}
	 {\setlength{\leftmargin}{10 pt}}}
\newcommand{\li}{\end{list}}

\newcommand{\Sum}{\displaystyle \sum}
\newcommand{\hi}{\bf}

\begin{document}

\title{Testing for financial crashes using the Log Periodic Power Law model}

%
%

\maketitle

{\bf Abstract}

Many papers claim that a Log Periodic Power Law (LPPL)  model fitted to financial market bubbles that precede large market falls or `crashes', contain parameters that are confined within certain ranges.
Further, it is claimed that the underlying model is based on influence percolation and a martingale condition.
This paper examines these claims  and their validity for capturing  large price falls in the Hang Seng stock market index over the period 1970 to 2008. The fitted LPPLs have parameter values within the ranges specified {\em post hoc} by \cite{JS:2001} for only seven of these 11 crashes. Interestingly, the LPPL fit could have predicted the substantial fall in the Hang Seng index during the recent global downturn. Overall, the mechanism posited as underlying the LPPL model does not do so, and the data used to support the fit of the LPPL model to bubbles does so only partially. 



{\it Classification codes:} G17, G01, C46.

\vspace{2 in}
{\footnotesize
We acknowledge the great help provided by Anders Johansen particularly in answering the many queries that we had about his data analysis and in providing us with the program that he used to estimate the parameters of the LPPL.}
\newpage

\section{Introduction}
\label{s:intro}

Financial crashes and the bubbles associated with them have generated much research interest particularly during the last few years. Since during  bubbles prices appear to move  away from the fundamental value of securities, financial bubbles are often attributed to the irrational exuberance of investors. If past earnings are high and investors use past earnings to predict future earnings, then it is likely that prices can contain long memory leading to stock price bubbles. However, the evidence  for long memory in financial prices is essentially mixed (see, e.g., \cite{AAM:1993, HRS:2010}). Similarly, financial bubbles seem much more substantial than one would expect. The price falls and  volatility that follow when prices are unsustainably high, lead to prices that are below  fundamental value. Such extreme movements in prices have important implications for risk management and valuation of financial securities.  Not surprisingly, there is ongoing debate in both the academic and business press on the extent to  which financial regulation can curb such extreme price movements.  

Two broad finance theories make predictions about stock price changes. They are the efficient market hypothesis (EMH) and the rational bubbles view (RBV). Both theories begin from the standpoint that an asset has a fundamental value, defined as the market's expected discounted present value of the firm's future cash flow that impacts on the value of the firm's  stock price. Empirical tests of both the EMH and the RBV often fail to explain large market price falls or `crashes', since such financial crashes are not usually associated with any specific news item.\footnote{
Our definition of a stock market crash is similar to that of  \cite{HS:2003}, in that they represent unusual large market falls that are not followed by large public news events and where such falls are market wide in nature. This definition accords with certain empirical work.} 
For example,  \cite{CPS:1989} find that of the 50 largest daily price falls in aggregate stock prices for the period 1946-1987, the majority are not accompanied by external news of specific importance.\footnote{
Recently, several specific theoretical models of stock market crashes have been put forward. 
In \cite{R:2001} symmetric rational asset-price model, neither rational behavior nor external news plays an important part in giving rise to stock market crashes. 
Both the  \cite{HS:2003} and \cite{BV:2003} models assume that economically significant differences in the views of investors can lead to stock market crashes when they are revealed.}  However, recent empirical work shows that external news might have a role to play in giving rise to financial crashes. Indeed, in a related study, Rangel (2011) finds that it is the surprise element of macroeconomic announcements that gives rise to significant jumps and volatility intensity, but only in respect of  inflation shocks represented in the Producer Price Index.

Empirical tests of the RBV have also had limited success in identifying price bubbles prior to large price falls (see \cite{BW:1982} and \cite{W:1987}). 
\cite{DK:1996} estimate  a non-linear ARMA-ARCH artificial neural network model that enables them to reject the claim that the 1929 stock market crash was the outcome of a bubble. One reason for the failure of tests of the RBV is the difficulty of explicitly isolating an asset's fundamental value from the component of the bubble tied to the asset's market price. Recently,  \cite{K:2010a} put forward a theoretical explanation for the origin of bubbles, their persistence and the reason for the crash that follows. \cite{K:2010b} demonstrates that financial crashes originate primarily from the herding behavior of noise traders and the increase/decrease in the associated momentum in noise trading during the life cycle of noise trading activity.      

Empirical researchers employ a variety of approaches to model financial crashes or unusual price movements. \cite{Bal:2007} developed a conditional extreme value theory (EVT) in terms of value at risk (VaR)  that appears to outperform traditional approaches that rely on the skewed {\it t} or normal distribution for modeling unusual price movements (see also \cite{HMC:2006}). However, the distributional form of EVT based on VaR does not lead to a semi-martingale process so that, when seeking to  predict stock price movements, the model seems  inappropriate.   \cite{KRBMF:2011} employ several different distributional assumptions for the conditional errors of  their ARMA(1,1)-GARCH(1,1) model. They find that the predictive ability of their ARMA(1,1)-GARCH(1,1) model depends on the distributional assumptions that underly the conditional errors. Indeed, versions of the ARMA(1,1)-GARCH(1,1) model which assume  non-normally  distributed conditional errors  performed better than those that assume  a normal  distribution. Specifically, their classical tampered stable distribution ARMA(1,1)-GARCH(1,1) model performs best, providing early warning signals of up to one year of financial crashes, including the September 29, 2008 crash.  \cite{KMP:2003} apply logit models to both microeconomic and financial data and show that currency crashes can be predicted. Similarly, \cite{MKD:2009} use  an ordered logit regression to predict financial crashes. Their results show  that the global crashes tend to follow local and regional crashes in which interest rates and market volatility play an important part. 

In this study we employ an alternative approach to model  the financial  bubbles prior to crashes. We fit a Log-Periodic Power-Law (LPPL), due to  \cite{SJB:1996}, to the Hang Seng index . The LPPL  approach has attracted a lot of  attention in recent years. An important advantage of the LPPL model relative to other approaches is that  it seeks to predict both the continuation and termination of a bubble in the same estimation.
The notion that financial crashes are manifestations of power law accelerations essentially suggests that endogenously induced stock market crashes might obey a particular power law, with log-periodic fluctuations.  This approach contrasts with \cite{R:2011} where crashes are considered to be exogenously induced. Following \cite{SJB:1996}, 
and \cite{LM:2004}, many papers claim to show that this model is able to capture a shift over time in the log-periodic oscillations of financial prices that are associated with market crashes. 
Given the manner in which the estimation is performed, shifts over time log-periodic oscillations are not directly captured but depend on the window that is searched.

The analogy of financial crashes as being similar in their statistical signatures to critical points as depicted in natural phenomena has, however, been argued to be unrealistic. \cite{LPC:1999} express doubts about the validity of fitting a seven-parameter model  to highly noisy data. They argue that such a model would suffer from severe over-fitting. Also, some log-periodic precursors do not always lead to crashes but to a smooth draw-down or  even a greater draw-up. This suggests that there is no universal  manner in which financial bubbles  manifest themselves. Indeed, some evidence (see, e.g.,  \cite{F:2001}) shows that the predicted time of a crash is sensitive to the size of the event-window used to predict the crash. We also find the size of the event-window places an important constraint on our empirical results.

Whilst the LPPL model is not perfect, it is empirically appealing as it provides a forecast of the date by which a financial crash might occur.\footnote{
\cite[p. 4]{LPC:1999} report two instances when financial crashes were predicted ex ante. 
The prediction was correct in one case but not in the other despite both predictions being published prior to the expected crash date. 
Indeed, they conclude that ``\ldots recent claims on the predictability of crashes are at this point not trust worthy."}
 This is an important attribute relative to other methods of financial risk assessment. For example, \cite[p. 461]{NB:2006} argue that EVT provides a means of predicting ``\ldots the magnitude of a market crash but not the day of the event." Furthermore, the LPPL model contains a component that captures the market's excessive volatility prior to a crash. This feature is consistent with several theoretical models of financial crashes as well as with empirical results \cite{L:2008,C:1996}. Indeed, \cite{KK:2004} show that the tail of the cumulative distribution function of ensembles of changes in stock prices is well described by a power-law distribution. As such,  the LPPL model provides a reasonably good fit to the data (see also,  \cite{K:2006}). Overall, the LPPL model appears to contain important statistical attributes that require serious empirical consideration and we explore some of those features in this study.

There are several critical considerations associated with fitting an LPPL model to financial data: 
first, studies that support the LPPL model (see e.g., \cite{JLS:2000}) show that the parameter estimates of the LPPL model are confined within certain ranges and that it is these ranges that are the indicators of market crashes.
This approach considerably restricts the number of classes of permissible LPPL fits to just those fits with parameters that fall within the specified ranges rather than to LPPLs with any values for their seven parameters. 
This means that the choice of the parameters for determining a crash does not rely on some p-value; 
this is an important weakness in using the LPPL to identify financial crashes.

Second, the mechanism underlying the LPPL model is such that prices must be expected to increase throughout the bubble, which is largely in line with the rational bubbles literature, but which is not what has been found in early empirical fits of the LPPL model (see Section \ref{ss:underlying mechanism}). 
Finally, there has  been neither sufficient critical analysis of the LPPL model nor sufficient assessment of its goodness-of-fit to available data. In particular, a goodness-of-fit test is rarely applied in empirical work and the sensitivity of the parameters of the fitted LPPL model is usually not reported (see Section \ref{ss:sensitivity}).

The remaining main sections of this paper are as follows: Section~\ref{s:LPPL} introduces the LPPL model; Section~\ref{s:mechanism} describes the mechanism underlying the LPPL model and evaluates prior work; Section~\ref{s:fitting} gives some details of the procedure used for identifying  the parameters of an LPPL that  best fits the data; Section~\ref{s:results} presents the fits obtained for the  pre-crash bubbles of the Hang Seng index, compares  the parameters obtained with those of prior work as well as  tests whether the parameters obtained  have values that do in fact predict their following crashes. We provide a summary of our results and conclude in the last section.

\section{The LPPL}
\label{s:LPPL}

The simplest form of the LPPL model can be written as:

\begin{equation}
\label{lppl} y_t = A +B(t_c-t)^\beta\left\{1 + C\cos(\omega \log(t_c-t) + \phi)\right\}, 
\end{equation}

\noindent
where:\\
\begin{tabular}{ll}
$y_t > 0$	&is the price (index), or the log of the price, at time $t$;	\\ 
$A > 0$	&is the value that $y_t$ would have if the bubble were to last until\\ &the critical time $t_c$;	\\
$B < 0$	&is the decrease in $y_t$ over the time unit before the crash 
		if C is \\
		&close to zero 	\\
$C$		&is the magnitude of the fluctuations around the exponential \\
		&growth, as a proportion;	\\
$t_c>0$	&is the critical time;	\\
$t < t_c$	&is any time into the bubble, preceding $t_c$;	\\
$\beta= 0.33 \pm 0.18$		&is the exponent of the power law growth;	\\
$\omega = 6.36 \pm 1.56$	&is the frequency of the fluctuations during the bubble;	\\
$0\le \phi \le 2\pi$			&is a shift parameter.
\end{tabular}

\noindent The ranges of values given for both $\beta$ and $\omega$ are based on the observed parameters of crashes for many stock markets \cite{J:2003}. 
Researchers tend to rely on established ranges for $\beta$ and $\omega$, rather than any goodness-of-fit test, to identify the bubbles that precede crashes.

Empirical studies that fit the LPPL model to financial data make a number of claims:
\begin{enumerate}
\item	The mechanism that characterizes traders on financial markets is one in which they mutually influence each other within local neighborhoods. 
	This leads, in turn to coordinated behavior through a martingale condition, which in the extreme can lead to a bubble and then a crash (see e.g., \cite{JLS:2000}). 
\item Endogenously induced financial crashes are preceded by bubbles with fluctuations. Both the bubble and the crash can be captured by the LPPL model when specific bounds are imposed on the critical parameters $\beta$ and $\omega$  (see e.g., \cite{J:2003, JS:2001}).
\item The values of the parameters $\alpha$ and $\omega$ 
	for the empirically fitted LPPL are sufficient to distinguish between LPPL fits 
	that precede a crash from those that do not  (see e.g., \cite{SJ:2001}).\footnote{
		 \cite{LRS:2009} carried out such an evaluation on a variant of the LPPL model.}
\end{enumerate}
\noindent In this paper, we examine the first two of the above claims and suggest a new approach for testing them. The third claim is more controversial; it only makes sense to evaluate it once we have a positive evaluation of the second claim.

\section{Is the Underlying Mechanism Correct?}
\label{s:mechanism}

\subsection{The underlying mechanism}

The mechanism driving the change in price during a bubble as posited in \cite{JLS:2000} is based on rational expectations, 
namely, that the expected price rise must compensate for the expected risk.
The mechanism is a stochastic process such that the conditional expected value of the asset at time $t+1$, given all previous data before and up to $t$, is equal to its price at time $t$. The martingale condition as formulated by \cite{JLS:2000} is:
\begin{equation}
	\label{dp1}	dp \leftarrow \kappa p(t) h(t) dt,
\end{equation}

\noindent
\begin{tabular}{lll}
where:	&$dp$	&is the expected change in price, conditional on no crash occurring \\
		&		& over the next time interval $dt$, at equilibrium;	\\
		&$p(t)$	& is the price at time $t$;	\\
		&$\kappa$&is the proportion by which the price is expected to drop during \\
		&		& a crash, if it were to occur; \\
		&$h(t)$	&is the hazard rate at time $t$, i.e. the chance the crash will occur \\
		&		& in the next unit of time,	given that it has not occurred already.	\\
\end{tabular}

\noindent 
Under this martingale condition, investors will buy shares at time $t$ if they expect the price at time $t+1$ will exceed the price at $t$ by more than the associated risk. That is: $ E(p(t+1)) > p(t) + dp$. This buying would drive up today's price.
So the expected rise in price between today and tomorrow will be less (assuming that the expected price tomorrow remains constant); this buying will continue until the expected rise is in line with the perceived risk according to Eq. \ref{dp1}. 
Alternatively, if investors believe that the expected rise in price tomorrow will be insufficient to compensate for the risk, i.e. $ E(p(t+1)) < p(t) + dp$, then they will  sell today, going short if necessary, thus driving today's price down.

Notice that all the terms on the right side of Eq. \ref{dp1} are positive, so $dp > 0$, 
i.e., the price must always be expected to be increasing during a bubble. 
This condition  was not treated as a constraint in early work (see, e.g., \cite{JLS:2000}) and as such gives us the opportunity of treating this requirement as a testable prediction.\footnote{
	 \cite{SZ:2006} does treat this condition as a constraint on the permissible parameter values.}

We now follow the consequences of Eq. \ref{dp1} for the behavior of prices.
Re-arranging Eq. \ref{dp1} gives us:
\begin{eqnarray}
\nonumber	\frac{1}{p(t)} dp &=& \kappa h(t) dt,		\\
\label{p=h} 	\log p(t) 	&=& \kappa \int_{t_0}^t	h(t') dt' .	
\end{eqnarray}

To capture the behavior of the price, the hazard rate, $h(t)$, needs to be specified.
Here, \cite{JLS:2000} posit a model in which each trader $i$ is in one of two states, either bull (+1) or bear (-1). At the next time step, the position of trader $i$ is given by:

\begin{equation}
\label{trader}
	\mathrm{sign} \left( K \Sum_{j \in N(i)} s_j + \sigma \epsilon_i \right),
\end{equation}
\noindent 
\begin{tabular}{lll}
where:	&$K$ 	&is the coupling strength between traders;	\\
		&$N(i)$ 	&is the set of traders who influence trader $i$;	\\
		&$s_j$	&is the current state of trader $j$;	\\
		&$\sigma$ 	&is the tendency towards idiosyncratic behavior for all traders;	\\
		&$\epsilon_i$ 	&is a random draw from a normal distribution with zero mean \\
		&			&unit variance.  \\
\end{tabular}

\noindent 
The relevant parameter determining the behavior of a collection of such traders is the ratio $K/\sigma$, which determines a critical value of $K$, say $K_c$.
If $K\ll K_c$ then the collection is in a disordered state. 
However, as $K$ approaches $K_c$ order begins to appear in the collection, with a majority of traders having the same state.
As the value of $K$ approaches $K_c$ from below, the system becomes more sensitive to small initial perturbations. 
At the critical value, $K_c$, all the traders will have the same state, either +1 or -1.
\cite{JLS:2000} further assume that: i) the coupling strength of $K$ increases smoothly over time up to $K_c$; and ii) the hazard rate is proportional to $K$. They do not justify these assumptions but the first one might be based on assuming that, as the frequency of fluctuations increases, traders become less sure of their own judgment and rely more on the judgment of their neighbors. In the next sections, we consider the evolution of $K$ over time.

\subsection{Simple power law hazard rate }
In the simplest scenario, $K$ evolves linearly with time.
Assuming that each trader has four neighbors arranged in a regular two dimensional grid, then the {\em susceptibility} of the system near the critical value, $K_c$, can be shown to be given by the approximation:
\begin{equation}
	\chi \approx B'' (K_c-K)^{-\gamma},
\end{equation}
\noindent
where $B''>0$ and $0 < \gamma < 1$ (see \cite{JLS:2000}). 
The three assumptions taken together give:
\begin{equation}
	\label{h1}	h(t) \approx B' (t_c-t)^{-\alpha},
\end{equation}

\noindent where $0<\alpha<1$.
Substituting in Eq. \ref{p=h} for $h$ as given by Eq. \ref{h1} and integrating gives:
\begin{eqnarray}
\nonumber \log p(t)	&=& \kappa \int_{t_0}^t B'(t_c-t')^{-\alpha} dt'	
				= \frac{-\kappa B' }{1-\alpha} \left[(t_c-t)^{1-\alpha}\right]^t_{t_0}\\
\nonumber 		&=&\frac{-\kappa B' }{1-\alpha} \left((t_c-t)^{1-\alpha} - (t_c-t_0)^{1-\alpha}\right).\\
\nonumber \mbox{At } t= t_c,\ \log p(t_c)	&=&\frac{-\kappa B'} {1-\alpha} \left(0 - (t_c-t_0)^{1-\alpha}\right).	\\
\nonumber \mbox{So } \log p(t) 	&=& \log p(t_c) - \frac{\kappa B'}{1-\alpha} (t_c-t)^{1-\alpha} 	\\
\label{p1}	 				 	&=& A + B (t_c-t)^{\beta},
\end{eqnarray}
where:
 $A	= \log p(t_c),\ 
 B 	= - \kappa B'/(1-\alpha) $ and $\beta = 1-\alpha$.
This is a simple faster-than-exponential growth model.

\subsection{Log periodic hazard rate}
To introduce log periodic fluctuations into the growth function, we need a different form of interconnected structure. 
Such a structure is assumed to be equivalent to one created by: i) starting with a pair of linked traders; ii) replacing each link in the current network by a diamond with four links and two new nodes diagonally opposite each other. This process continues until some stopping criterion is met. Then (see \cite{JLS:2000}):
\begin{eqnarray}
\nonumber \chi 	&\approx 
		&B'' (K_c-K)^{-\gamma} + C''(K_c-K)^{-\gamma} \cos(\omega \log(K_c-K) + \phi') + \ldots .\\
\label{h2}	\mbox{So }	h(t)	&\approx 
		&B'(t_c-t)^{-\alpha}\{1 + C' \cos(\omega \log(t_c-t) + \phi')]\}
			\mbox{, from Eq. \ref{h1}.}
\end{eqnarray}

\noindent
Substituting for $h$ in Eq. \ref{p=h} from Eq. \ref{h2} and integrating gives 
(see appendix for details):

\begin{equation}
\label{p3}	  \log p(t) =
 			A + B(t_c-t)^{\beta}\left\{1+C\cos(\omega \log(t_c-t) + \phi)\right\},
\end{equation}

\noindent 
which is the LPPL of Eq. \ref{lppl} with $y_t = \log(p_t)$.

\subsection{Index: raw versus log} 
Note from  Eq. \ref{p3} that it is the log of the price index that needs to be fitted to the LPPL, although in practice the LPPL model has often been fitted to the raw index data.
\cite{JS:2001} recommend the use of the raw data 
when the price drop in the crash is proportional to the price over and above the fundamental value rather than being proportional simply to the price.
That is, they replace the condition \ref{dp1} by:
	
\begin{equation}
	\label{dp2}	dp \leftarrow \kappa (p(t)-p_1) h(t) dt,
\end{equation}
\noindent where $p_1$ is the fundamental value (which they do not further define).

\cite{JS:2001}  introduce the assumption that the rise in price since the beginning of the bubble is much less than the amount by which the price at the beginning of the bubble is above the fundamental value. Thus 
\begin{equation}
	\label{small price rise}	 p(t)-p(t_0) \ll p(t_0) -p_1, 
\end{equation}
where $t_0$ is the time of the beginning of the bubble.
Even if the asset's fundamental value is not estimated in the model, the above assumption is weakly testable.
If the price rise during the bubble is greater than the price at the beginning of the bubble, i.e. $p(t) > 2 p(t_0)$, then the condition of Eq. \ref{small price rise} cannot be fulfilled unless the fundamental price is negative.
We assume that this is not what is intended. 
So we can test whether or not this assumption is met.

Integrating Eq. \ref{dp2} from the moment when the bubble starts, $t_0$, and using Eq. \ref{small price rise} gives:
\begin{eqnarray}
\nonumber	p(t) &=& p(t_0) + \int^t_{t_0} dp \\
\nonumber	&=& 	p(t_0)+\kappa \int^t_{t_0} (p(t')-p_1) h(t') dt'	\\
\label{p4}		&\approx &	p(t_0) +\kappa (p(t_0) - p_1) \int^t_{t_0} h(t') dt'.
\end{eqnarray}
Provided the assumption in Eq. \ref{small price rise} is met, Eq. \ref{p4} can be used to fit the LPPL to raw price  (as done, e.g., in \cite{JS:2001}) rather than the log price data.

\subsection{Tests of the underlying mechanism}
\label{ss:underlying mechanism}
\cite{CF:2006}  tested the mechanism underlying the LPPL model using S\&P index data for the bubble preceding the 1987 crash. 
They  compared the predictions of a LPPL fitted to the data with  a random walk model. 
To do so, they first extended the LPPL model as given in Eq. \ref{lppl}, by adding:
\LI
\item a random term with zero mean and variance estimated from the data.
	This noise term is necessary to compute a likelihood for the observed data deviations from the predicted LPPL model.
\item a positive upward drift term estimated from the data.
	This addition to the LPPL model, while frequently made in financial time series, is unnecessary here, as faster-than-exponential growth is posited in the LPPL model.
\li
Then they estimated the likelihood of the observed change in price since the previous day, $t-1$, and selected parameters that maximized the sum of these likelihoods over the entire bubble.

With a time series there is a choice of which next point to take as being the most likely: 
either the predicted value or the predicted change since $t-1$.
Using the model's prediction of the value at $t$ ignores the value at $t-1$; 
this is what \cite{JLS:2000} implicitly assume when they minimize the root mean square error for the fitted LPPL against the data.
On the other hand, using the predicted change since $t-1$ ignores any deviation that the price at $t-1$ already has, from the model's prediction for $t-1$. 
This is what \cite{CF:2006} explicitly do to
 specify the mechanism underlying their adaptation of the LPPL model. Not surprisingly, when judged for each time point separately, their method is not to be preferred to the random walk model \cite{CF:2006}.

While most of the assumptions underlying the mechanism from which the LPPL model is derived are untestable (or even questionable), there is one that is testable: the hazard rate $h$ must be positive.
This implies that the expected price must always rise.
If the fitted LPPL does not have this property, then the assumption that $h(t)$ in Eq. \ref{dp1} is a probability, must be rejected.

As proposed by \cite{GM:2003}, it is possible to force the hazard rate to be positive, . 
The condition for the hazard rate to be positive is, from Eq. \ref{h2}, that:
\begin{eqnarray}
\nonumber h(t) \ge 0 & \Leftrightarrow &
		B'(t_c-t)^{-\alpha}\{1 + C' \cos(\omega \log(t_c-t) + \phi')\} \ge 0 \\
\nonumber  & \Leftrightarrow &
		1 + C' \cos(\omega \log(t_c-t) + \phi') \ge 0 
			\mbox{, since } B'= -\beta B/ \kappa \ge 0 \mbox{ and } t_c \ge t\\
\nonumber  & \Leftrightarrow &
		|C'| \le 1\mbox{, since } |\cos| \le 1	Å\\
\label{slopePos}	  & \Leftrightarrow &
		|C| \le \beta/\sqrt{\beta^2+\omega^2} \mbox{, since } C =  \beta C'/\sqrt{\beta^2+\omega^2},
\end{eqnarray}
a condition that was used by \cite[equation 3]{SZ:2006}.
Requiring the slope of $y(t)$, as given in Eq. \ref{lppl}, to be positive,
i.e. $dy/dt \ge 0$, gives the same condition as Eq. \ref{slopePos}.
 \cite{GM:2003}, 
using a the three-year data window on the Dow Jones index between 1912 and 2000,
found that the condition \ref{slopePos} together with $7 < \omega < 13$, 
predicts that a crash would occur within a year 
on only a quarter (65/229) of the windows which were 
actually followed by a crash within a year.
So forcing the hazard rate to be positive here led to poor predictions.

\section{Fitting the LPPL Parameters}
\label{s:fitting}

The seven parameters of the LPPL in Eq. \ref{lppl} have to be estimated from the window of data points in the bubble. 
The chosen values of these parameters should be the ones that minimize the root mean squared error (RMSE) between the data and the LPPL model's prediction for each day of the bubble. 
The squared error between the prediction from the fitted curve from Eq. \ref{lppl} and the data is:

\begin{equation}
\label{rmse} SE = \Sum^{t_n}_{t=t_1} (y_t -\hat{y}_t)^2 
			= \Sum^{t_n}_{t=t_1} \left\{y_t - A -B(t_c-t)^\beta
					\left(1 + C\cos(\omega \log(t_c-t) + \phi\right)\right\}^2,
\end{equation}

\noindent
\begin{tabular}{lll}
where: &$y_t $		&is the data point, either the price index or its log; \\ 
	&$\hat{y_t} $	&is the data point as predicted by the model;	\\
	&$n$		&is the number of weekdays in the bubble;	\\
	&$t_i$		&is the calendar day date of the $i^{th}$ weekday \\
	&			&from the beginning of the bubble.	\\
\end{tabular}

\noindent
Partially differentiating Eq. \ref{rmse} with respect to the parameters $A, B$ and $C$ gives us three linear equations from which the values of $A, B$ and $C$ that minimize the RMSE are derived, given the other four parameters: $\beta, \omega, t_c$ and $\phi$. 
To find suitable values for these four parameters a search method is required. 
This search method used in \cite{JS:2001} and \cite{SJ:2001}, hereafter collectively called the {\em JS studies}, was:
\LI
\item First to make a grid of points for the parameters $\omega$ and $t_c$, from each of which a Taboo search was conducted to find the best value of $\beta$ and $\phi$, i.e. the ones for which, with $A, B$ and $C$ chosen to minimize the RMSE, gave the lowest RMSE.
\item To select from these points those for which $0<\beta<1$.
\item From these points, i.e. those points that were found to minimize the RMSE for which $0<\beta<1$,   conduct a \cite{NM:1965} Simplex search, with all the four search parameters free (and $A, B$ and $C$ chosen to minimize the RMSE).
\li
\noindent We presume that the reason that any fit with $\beta\ge 1$ was rejected is because  the increase in the index is exponentially {\em declining} whereas the underlying mechanism requires it to be increasing. An alternative technique would have been to place no restriction on the value of $\beta$, and if a value of $\beta \ge 1$ is found, to reject the model, as we have done for the requirement that the fitted LPPL never decreases (see Section \ref{ss:underlying mechanism test}). 

Similar to the JS studies, we use a preliminary search procedure based on a grid to provide seeds for the Nelder-Mead Simplex method, as implemented in Matlab \cite{LRWW:1998}.
It is based on choosing different values for the two parameters $\omega$ and $\beta$, as these are the critical parameters for determining whether the fitted LPPL model is a crash precursor or not (see Eq. \ref{lppl}).
The algorithm and the parameter values used are shown in the Appendix.
Note that instead of the crash date, $t_c$, we use $t2c$, the number of days between the day on which the estimate is being made and the predicted critical date.

\section{Empirical Results}
\label{s:results}

\subsection{Test of the underlying mechanism}
\label{ss:underlying mechanism test}

\begin{figure}[t]
\centering
\resizebox{\textwidth}{!}{\includegraphics{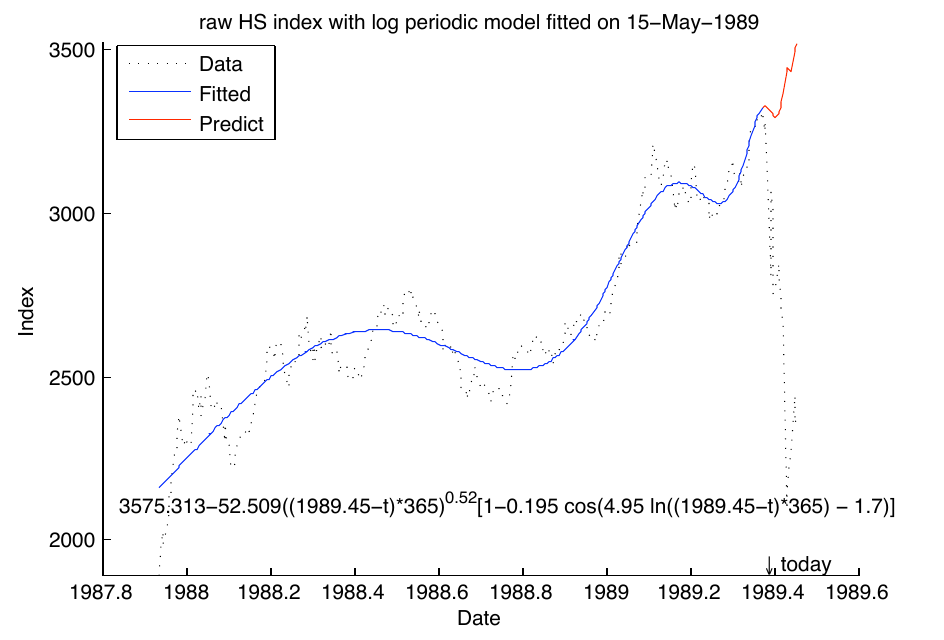}}
\caption{LPPL fit to the bubble preceding the 1989 crash on Hang Seng.}
\label{fig:HS1989}
\end{figure}

In this section we test whether or not the underlying mechanism,  as described in 
Section~\ref{ss:underlying mechanism}, applies to the raw Hang Seng index data.
The observations for the Hang Seng index were obtained from Datastream. We analyze the Hang Seng index since it is commonly believed that this stock market has had several crashes, thus giving us ample opportunity to test the LPPL model.\footnote{ 
This suggests that stock market crashes can be common. Indeed, using a statistical method to identify outliers, \cite{ST:2008} show that the 1987 stock market crash of the Dow Jones Industrial index was not a structurally unusual event.}

As an initial test, we  show the LPPL fitted to the raw Hang Seng index data for the bubble preceding the 1989 crash. 
We use this crash period for the Hang Seng index in order to closely match this part of our results with those of  \cite{SJ:2001}. 
The plots of the LPPL model are shown   in Figure~\ref{fig:HS1989}. 
The fit of our LPPL model is similar to Figure 8 of  \cite{SJ:2001}. Notice that the LPPL in Figure 1 has a negative slope some of the time. 
The same is true in 18 of the 30 cases reported in  \cite{JS:2001} and \cite{SJ:2001}.\footnote{
	These  16 pre-crash  bubbles are: the Dow Jones (1929, '62), S \& P ('37, '87), Hang Seng ('80, '89, '94, '97), 
	Argentina ('91, '92, '97) and various other stock market crashes of 1994 
	(Indonesia, Korea, Malaysia, Philippines) and 1997 (Indonesia, Mexico, Peru).} 
That is, the fitted LPPL predicts that  on  average the price should decrease at some time points.
This empirical fact is sufficient to reject the martingale condition as being the mechanism underlying the LPPL fit to pre-crash bubbles.

\subsection{Data and descriptive statistics}

\begin{table}[ht]
\begin{center}
\begin{minipage}{11 cm}
\caption{Descriptive statistics for changes in the log of the Hang Seng stock index.}
\label{tab:Hang Seng stats}
\begin{tabular}{cccccc}
		N	&Mean	&Variance	&Skew	&Kurtosis	&\begin{tabular}{c}Jarque-Bera \\
																statistic\end{tabular}	\\
\hline
		10152	&0.00045$^b$	&0.00035	&-1.25934$^a$	&31.58011$^a$	&424542.78.9$^a$\\
\hline
\end{tabular}
~\\
\textit{Note}: The mean and variance are multiplied by 100\\
$^a$ denotes statistical significance at the 1 percent level\\
$^b$ denotes statistical significance at the 5 percent level
\end{minipage}
\end{center}

\end{table}

To perform more rigorous tests on the  fits of the LPPL model, we extend  the daily prices for the Hang Seng to cover the period 1$^{\mbox{\footnotesize st}}$ January 1970 to  31$^{\mbox{\footnotesize st}}$ December 2008. 
Descriptive statistics, shown in Table~\ref{tab:Hang Seng stats}, reveal that the mean log changes of the Hang Seng index series are significantly different from zero. Both skewness and (excess) kurtosis are significant such that the Jarque-Bera test rejects the null of normality at a 1 percent level. Notice that skewness is highly significant and negative. 
This finding suggests that the Hang Seng stock market can be very sensitive to stock market crashes. That is, volatility feedback can increase the probability of large negative returns and in turn, increase the potential for crashes \cite{CH:1992}.

\subsection{Identifying a crash}
\label{ss:what is a crash}

To test whether or not the LPPL can predict crashes we first need to identify the crash itself. 
Usually a stock market crash is taken to mean a very large and unusual price fall. In our application, a crash can span more than one day. This is consistent with the October 1987 stock market crash. 

\begin{figure}[t]
\centering
\resizebox{\textwidth}{9 cm}{\includegraphics{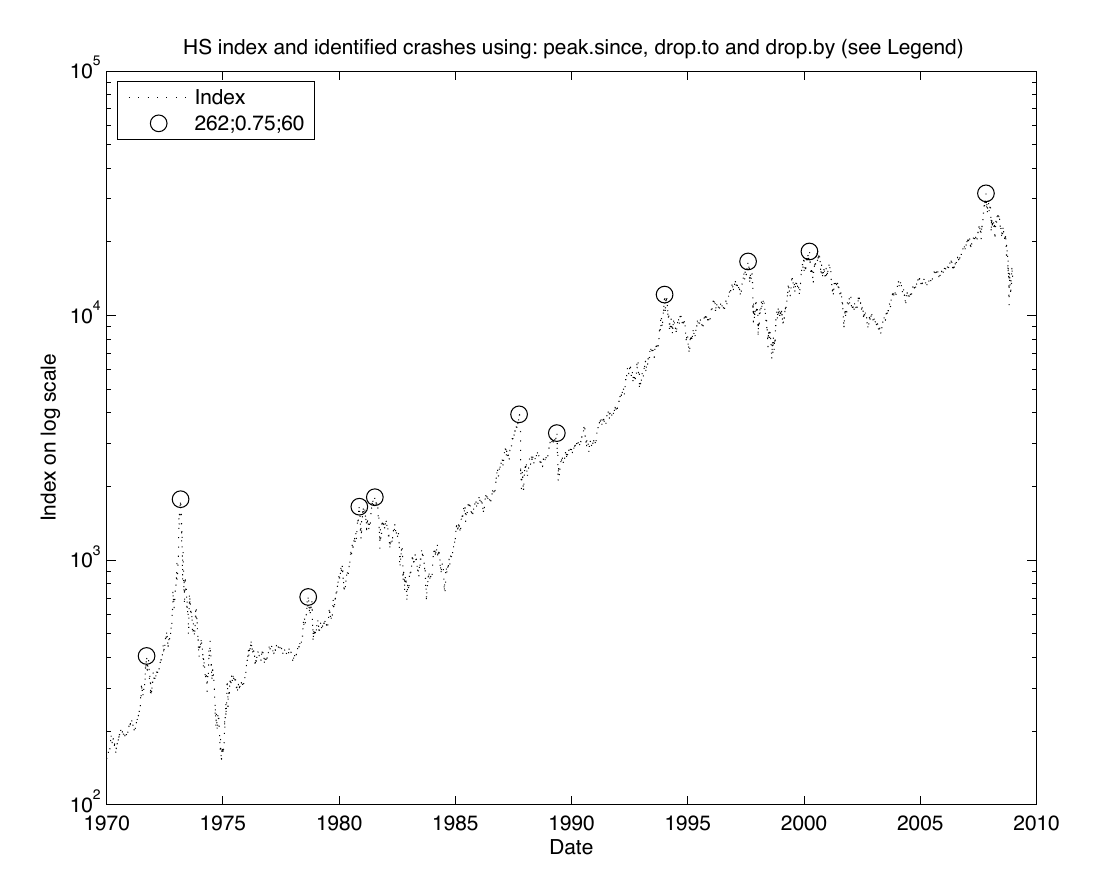}}
\caption{The Hang Seng index 1970 to 2008, showing those peaks that are initiators of crashes.}
\label{fig:Hang Seng_index}
\end{figure}

There are two situations when we might falsely claim that a crash has occurred. 
One is when the index is on the way up in a bubble and then there is a large drop, but it turns out that the drop is temporary and the bubble continues. 
The other is when, on the way down during a crash, the index experiences a recovery and so we identify the beginning of a new bubble but the recovery is temporary and the anti bubble is still in effect. 
To avoid those situations, we identify a peak as one initiating a crash as follows:
\LI
\item	a period of 262 weekdays prior to the peak for which there is no value higher than the peak,
	 
\item	a drop in price of 25\%, i.e. down to 0.75 of the peak price, which is in line with the 1987 crash,

\item	a period of 60 weekdays within which the drop in price needs to occur.
		
\li

\begin{figure}
\centering
\resizebox{\textwidth}{9 cm}{\includegraphics{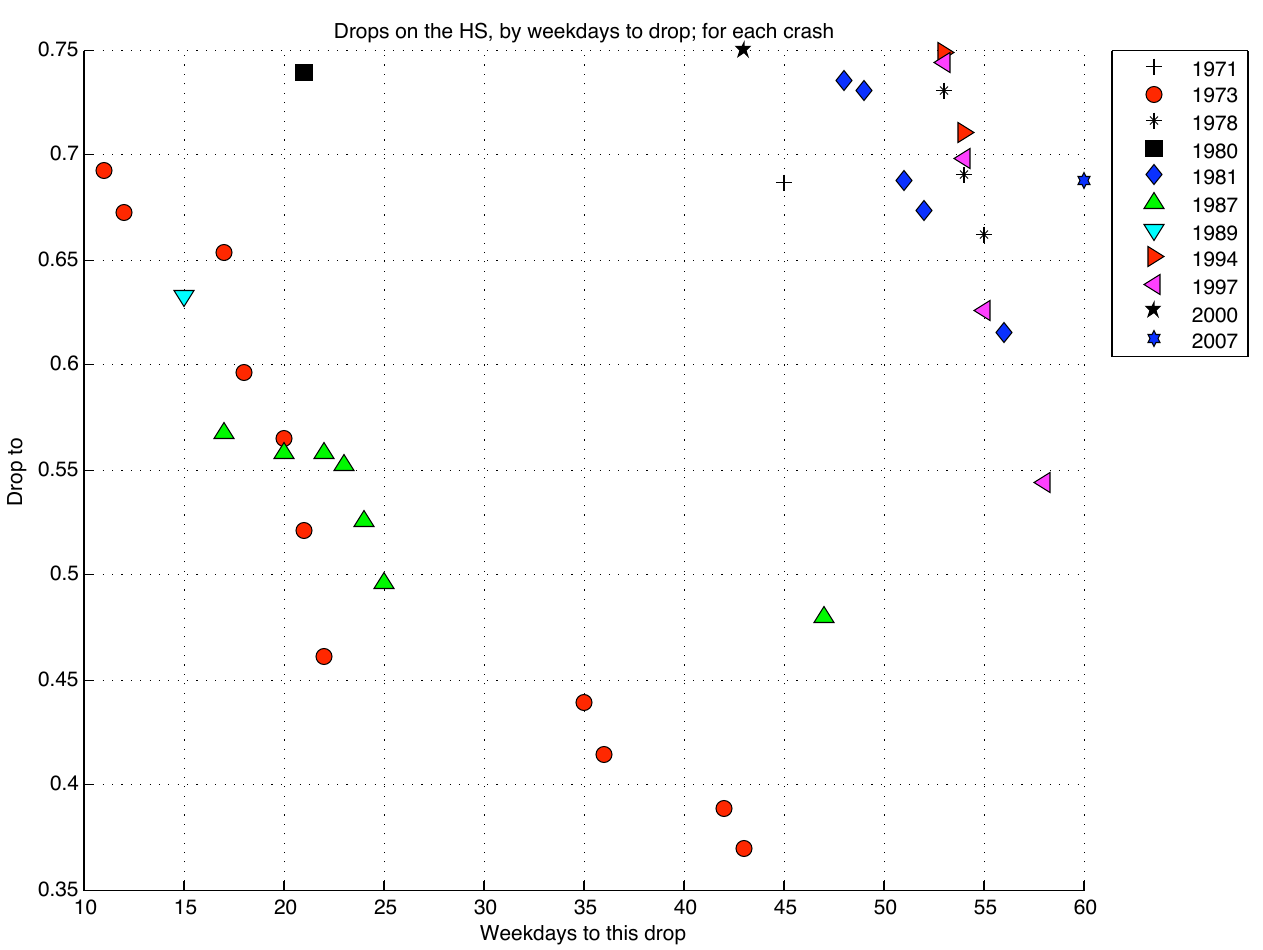}}
\caption{Drops from peaks on Hang Seng index 1970 to 2008.}
\label{fig:Hang Seng_drops}
\end{figure}

We first tested whether the application of these criteria enables us to capture the eight crashes on the Hang Seng index, as identified in the JS studies.
Indeed, we identify crashes at the same time points as in the JS studies, except for one additional crash in 1981 (see Figure~\ref{fig:Hang Seng_index}). To exclude the price fall in 1981 from being classified as a crash, we would have to increase the drop-to criterion or reduce the drop-by criterion.
Doing either would also exclude some of the other peaks as initiating crashes, viz. those peaks that immediately preceded the crashes of 1978, 1994, 1997, all of which are identified as crash initiators in the JS studies (see Figure~\ref{fig:Hang Seng_drops}). Thus the rule they apply seems somewhat imprecise.
It is true that the 1981 crash occurs shortly after the 1980 crash, so we might exclude the 1980 peak as initiating a crash, but rather being a part of the bubble preceding the 1981 crash, but this is not what was done in  \cite{SJ:2001}.
It would also be possible to exclude fitting an LPPL model to the bubble preceding the 1981 crash on the grounds that this bubble is too short -- just 7 months long. 
However, another bubble (the one preceding the crash 1971) was fitted even though it lasted only 6 months.
As such, the bubble preceding the 1981 crash should have been included in the JS studies, unless one insists on having more than say 7 months of data preceding a crash. 
On balance, we believe that it is appropriate to include the 1981 crash we have identified, giving us nine crashes for the period of the JS studies.  Overall, the criteria for identifying a crash does not appear to be consistently applied in the JS studies.

In the period after the JS studies, i.e. between 2000 and 2008, our criteria identify two additional peaks as initiating crashes; these are in 2000 and in 2007. The two bubbles preceding these crashes provide a post-hoc test of the hypothesis underlying the LPPL model (see Eq. \ref{lppl}).

\subsection{Troughs and bubble beginnings}
\label{ss:bubble beginnings}
Having decided that a peak is the initiator of a crash, the data window  to be used for fitting the LPPL model to the preceding bubble needs to be carefully selected.
In the JS studies the start of the data window is taken to be the day on which the index reaches its lowest value ``prior to the change in trend" \cite{JS:2001}. 
In real time matters are not so simple, since one does not know if the index will drop still further in the future. 
So for real time analysis we would need to take as the end of the previous crash the lowest point since the last crash, up until now. 

\begin{figure} 
\centering
\resizebox{\textwidth}{9 cm}{\includegraphics{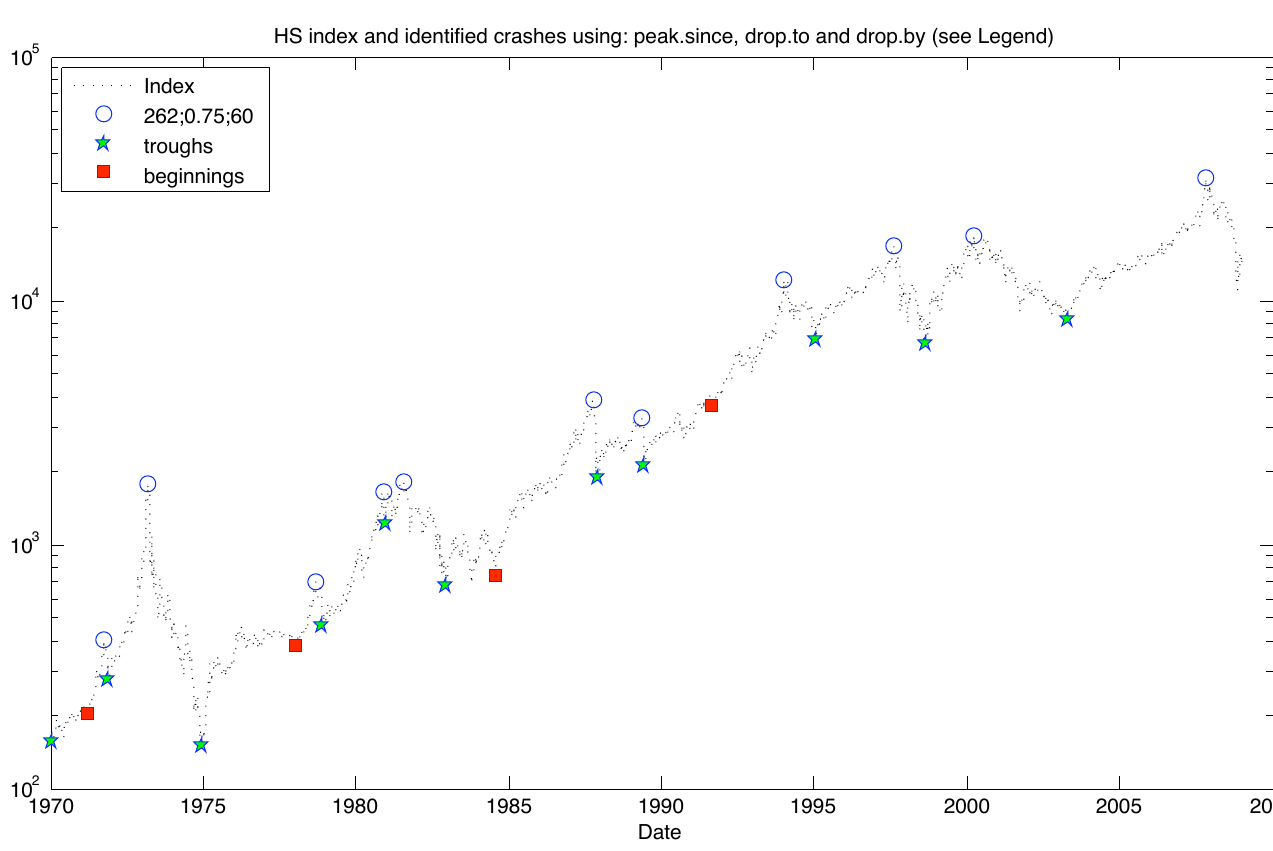}}
\caption{Troughs and other beginnings of bubbles on Hang Seng 1970 to 2008.}
\label{fig:Hang Seng_troughs}
\end{figure}

Moreover,  \cite{JS:2001} sometimes move the beginning of the bubble from the lowest point since the previous crash to a later time as in their Asian and Latin-American study. This was done if ``at the trough the next bubble had not yet begun" (Johansen, personal communication).
From the JS studies, we deduce that this was done for four of the eight crashes
they identified on the Hang Seng:
\LI
\item 1971 crash: forward 2 months, from 5/1/1971 to 10/3/1971,
\item 1978 crash: forward 3 years and 1 month, from 10/12/1974 to 13/1/1978,
\item 1987 crash: forward 1 year and 8 months, from 2/12/1982 to 23/7/1984,
\item 1994 crash: forward 2 years and 2 months, from 5/6/1989 to 19/8/1991.
\li
\noindent
These are indicated by squares in Figure \ref{fig:Hang Seng_troughs}.

It is clear why \cite{JS:2001} moved the beginning of the bubbles for the 1978 and 1987 crashes to times later than the trough proceeding the crash. 
For 1978 there was a long period of stable prices which is clearly not part of a bubble. 
For 1987, the year and 8 months following the trough are characterized by two mini bubbles and two peaks (which with other crash criteria would themselves be considered initiators of crashes). 
It is not so clear why they moved the start points of the other two bubbles (preceding the 1971 and 1994 crashes) forward. 

In the JS studies, a model fit is only made if there are at least 131 weekdays of data between the trough and the crash. Changing the number of days could lead to different bubbles being considered as crash precursors. To illustrate this for the Hang Seng data, there are only 155 weekdays between the end of the 1980 crash and the peak in 1981 when it appears that another crash occurred. To require (say) 262 weekdays would result in insufficient data, and thus exclude the bubbles before both the 1981 and the 1971 crashes, thus affecting the results. This means that one needs to be very careful in implementing the rule, given the data under consideration.

\subsection{Fitting to the raw index}
\label{ss:index}

\begin{table}[ht]
\begin{center}
\begin{minipage}{10 cm}
\caption{Ratio of raw Hang Seng index on the last day to index at the beginning of the bubble.}
\label{tab:Hang Seng_raw_assumption}

\begin{tabular}{rr|rr|r}
\multicolumn{2}{c|}{Bubble:} &\multicolumn{2}{c|}{Raw Hang Seng:}	&Ratio:	\\
beginning at  $t_0$	&~~~~ending on  $t_e$			&$p(t_0)$	&$p(t_e)$	&$p(t_e)/p(t_0)$\\
\hline
*10-Mar-1971	&20-Sep-1971	& 201	& 406	&2.02$^{\dagger}$	\\
22-Nov-1971	&09-Mar-1973	& 279	& 1775	&6.36$^{\dagger}$	\\
*13-Jan-1978	&04-Sep-1978	& 383	& 707	&1.85	\\
20-Nov-1978	&13-Nov-1980	& 468	& 1655	&3.54$^{\dagger}$	\\
12-Dec-1980	&17-Jul-1981	&1222	& 1810	&1.48	\\
*23-Jul-1984	&01-Oct-1987	& 747	& 3950	&5.29$^{\dagger}$	\\
07-Dec-1987	&15-May-1989	&1895	& 3310	&1.75	\\
*19-Aug-1991	&04-Jan-1994 	&3723	&12201	&3.28$^{\dagger}$	\\
23-Jan-1995	&07-Aug-1997	&6968	&16673	&2.39$^{\dagger}$	\\
13-Aug-1998	&28-Mar-2000	&6660	&18302	&2.75$^{\dagger}$	\\
23-Apr-2003	&30-Oct-2007	&8520	&31638	&3.71$^{\dagger}$	\\
\hline			
\end{tabular}
~\\
\textit{Note}: $t_0$,	the day the bubble began;
			$t_e$,	the last day of the bubble\\
*	Bubble beginning  moved to later than the trough between peaks\\
$^{\dagger}$	$p(t_e)/p(t_0)> 2$, so 
		the raw index should not be used\\
\end{minipage}
\end{center}
\end{table}

In the JS studies, for all but the 1973 crash, the LPPL model has been fitted to the bubble in the raw index rather than to the log of the index. 
For this to be justified, the inequality in Eq. \ref{small price rise} must hold. That is, the price rise during the bubble must be considerably less than the difference between the price at the beginning of the bubble and the fundamental price.
If we make the reasonable assumption that the fundamental price cannot be negative, then at any time during the bubble the expected price must at the very least not be more than double that at the beginning of the bubble. 
This condition is met for only two of the eight bubbles found in the JS studies (see Table \ref{tab:Hang Seng_raw_assumption}).
For the remaining six bubbles this condition does not hold, i.e. the expected price more than doubled during the bubble, so the inequality in Eq. \ref{small price rise}, which is the assumption upon which the raw rather than the log of the index can be chosen, was violated. Despite this, in the JS studies five of these six fits of the LPPL model are made to the raw index rather than to its log; they should not have been.

\subsection{Sensitivity to search parameter values}
\label{ss:sensitivity}

\begin{figure}[t]
\centering
\resizebox{\textwidth}{!}{\includegraphics{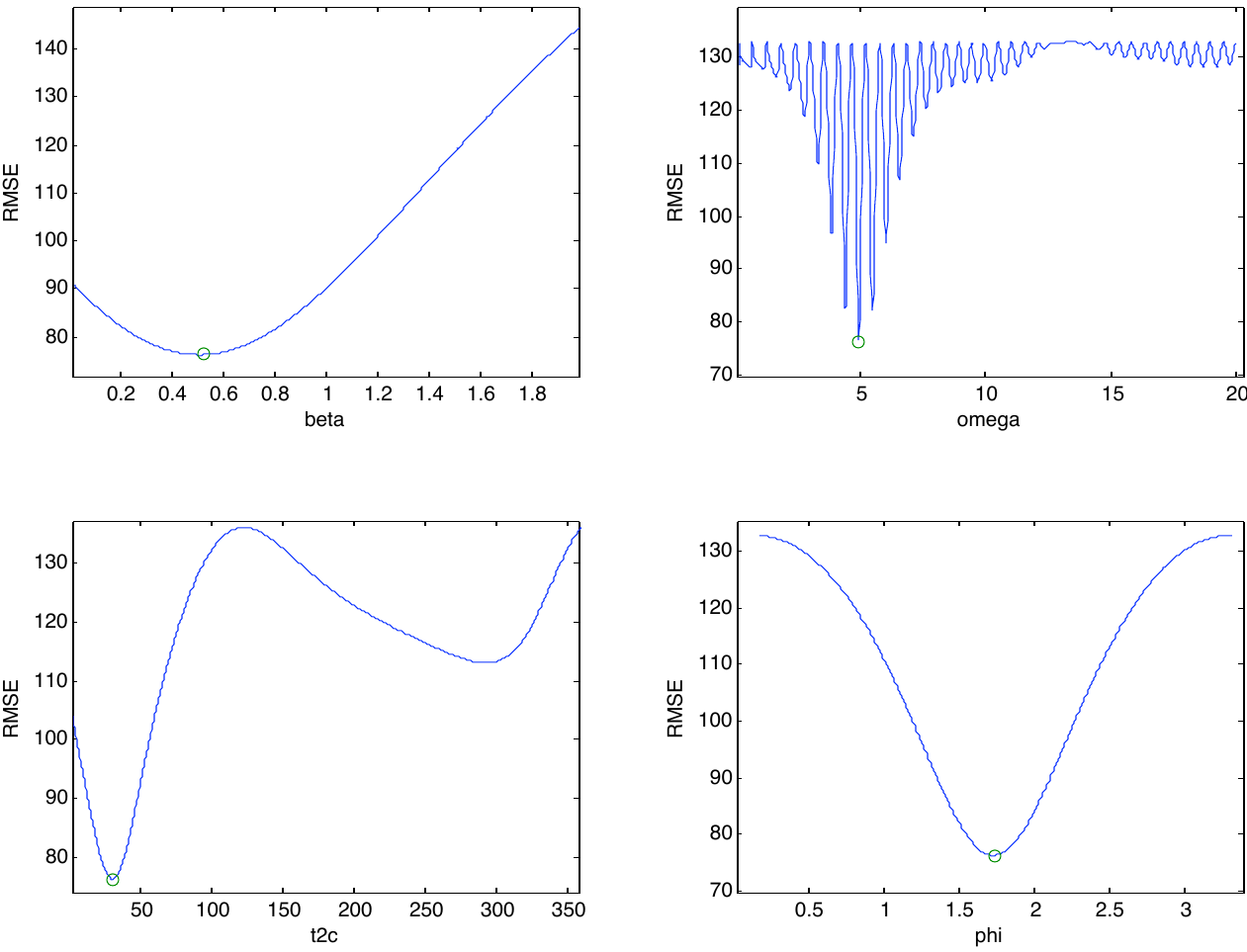}}
\caption{Sensitivity of the RMSE to the parameters of the LPPL for 1989 Hang Seng crash.}
\label{fig:sensitivity_1989_phi.jpg}
\end{figure}

Identifying an LPPL model fit to a bubble as one that precedes a crash depends on the values found for the two critical parameters $\beta$ and $\omega$; so it is important to examine how sensitive the RMSE of the fit is to variations in these parameters.
We use the bubble preceding the 1989 crash on the Hang Seng to examine the sensitivity of the  LPPL fit to variations in each of the four search parameters ($\beta, \omega, \phi$ and $t_c$); the other three parameters ($A, B$ and $C$) are always set using these four (see Section~\ref{s:fitting}). The results are shown in Figure~\ref{fig:sensitivity_1989_phi.jpg}. The circle indicates the chosen parameter value.
While the chosen values of the search parameters are at global minima, the RMSE is highly sensitive to small fluctuations in the value chosen for $\omega$ \cite{BCP:2010}.
The sensitivity diagrams for the other Hang Seng bubbles listed in Table~\ref{tab:Hang Seng_raw_assumption} are similar to those shown in Figure~\ref{fig:sensitivity_1989_phi.jpg}. 
Consequently the value found by the search procedure for $\omega$ may not be the one that leads to the minimum RMSE.
It might be thought that the search space could nevertheless be smooth: if a local minimum has been found, small variations in one or more of the other parameters might lead to a smooth surface and avoid the search procedure getting trapped. However, the sensitivity to other search parameters would then have to also fluctuate greatly, and they do not. 
So the multi-dimensional surface is unlikely to be smooth.
As the value of $\omega$ is used in predicting whether or not the bubble will be followed by a crash, this would be a serious problem.

\begin{table}
\begin{minipage}{\textwidth}
\caption{The bubbles and crashes of the Hang Seng index and LPPL fits to the raw bubble data.}
\label{tab:Hang Seng_fits}
{\footnotesize
\begin{tabular}{rr|rrr|r|r|rrr}
Parameter:	&			&A		&B		&C		&$\beta$	&$\omega$	&$t2c$	&$\phi$	&RMSE\\
Units:		&			&HSI		&HSI		&		&		&rads		&days	&rads	&HSI	\\
Predicted		&low:		&		&		&		&0.15	&4.80		&1		&0		&\\
range:		&high:		&		&0		&		&0.51	&7.92		&?		&$\pi$	&\\
\cline{1-2}
From/to		&Note		&		&		&		&		&			&		&		&\\
\hline
10-Mar-1971	&*[SJ]&594	&-132	&-0.033	&0.20	&4.30		&7		&0.50	&7.58	\\
20-Sep-{\bf 1971}&			&539	&-101	&-0.047	&0.22	&4.30		&3		&0.25	&6.11	\\
\hline
22-Nov-1971	&[SJ]	&11		&-3		&0.003	&0.11	&8.70		&2		&0.05	&0.0722	\\
09-Mar-{\bf 1973}&log		&65		&-56		&-0.001	&{\hi 0.01}	&{\hi 11.1}		&20		&1.32	&0.0538	\\
			&log			&8		&-0		&-0.177	&0.57	&{\hi 1.47}		&2		&3.14	&0.0549	\\	
			&raw			&2443	&-485	&-0.114	&0.26	&{\hi 1.45}		&2		&3.14	&40.91	\\
\hline
13-Jan-1978	&*[SJ]&816	&-50		&-0.053	&0.40	&5.90		&6		&0.17	&10.09	\\	
04-Sep-{\bf 1978}&			&741	&-23		&0.072	&0.51	&5.30		&1		&0.00	&10.12 	\\
\hline
20-Nov-1978	&[SJ]	&1998	&-231	&-0.044	&0.29	&7.24		&3		&1.80	&46.72	\\
13-Nov-{\bf 1980}&			&41164	&-38080	&0.001	&{\hi 0.01}	&7.51		&52		&3.06	&35.02	\\
			&			&7929	&-5352	&0.008	&{\hi 0.05}	&6.79		&26		&1.55	&35.55	\\
			&			&1998	&-231	&-0.044	&0.29	&7.24		&3		&2.63	&37.00	\\
\hline
12-Dec-1980	&--			&		&		&		&		&			&		&		&		\\
17-Jul-{\bf 1981}&			&1753	&-0		&-0.890	&{\hi 2.41}	&{\hi 3.02}		&1		&3.14	&40.46	\\
			&			&1817	&-3		&-0.567	&{\hi 1}	&4.75		&12		&0.35	&49.24	\\
			&			&1946	&-11		&-0.399	&{\hi 0.76}	&5.89		&36		&0.00	&54.95	\\

\hline
23-Jul-1984	&*[JS]&5262	&-542	&-0.007	&0.29	&5.60		&22		&1.60	&133.86	\\
01-Oct-{\bf 1987}&			&5779	&-711	&0.048	&0.27	&5.68		&34		&2.63	&68.47	\\
\hline
07-Dec-1987	&[SJ]	&3403	&-32		&-0.023	&0.57	&4.90		&34		&0.50	&133.21	\\
15-May-{\bf 1989}	&		&3575	&-53		&-0.195	&0.52	&4.95		&31		&1.74	&76.33	\\
\hline
19-Aug-1991	&*[JS]	&21421	&-7614	&0.024	&0.12	&6.30		&4		&0.60	&322.80	\\
04-Jan-{\bf 1994} &\multicolumn{2}{r}{212635}	&-194575 &-0.002	&0.27	&5.95		&1		&3.13	&272.82	\\
			&			&14038	&-1717	&-0.028	&0.26	&6.43		&4		&3.14	&281.36	\\
\hline
23-Jan-1995	&[JS]	&20359	&-1149	&-0.019	&0.34	&7.50		&51		&0.80	&531.79\\
07-Aug-{\bf 1997}	&		&20255	&-1201	&-0.048	&0.33	&7.47		&51		&2.29	&438.79	\\
\hline
13-Aug-1998	&--			&		&		&		&		&			&		&		&		\\
28-Mar-{\bf 2000}	&		&21918	&-16		&0.073	&{\hi 1.00}	&{\hi 18.35}	&290	&0.00	&710.99	\\
			&			&24095	&-97		&-0.057	&{\hi 0.76}	&{\hi 17.51}	&264	&3.14	&720.17	\\
			&			&19503	&-372	&0.111	&0.52 	&5.7 			&9 		&2.07	&744.15	\\
\hline			
23-Apr-2003	&--			&		&		&		&		&			&		&		&		\\
30-Oct-{\bf 2007}&			&38940	&-6408	&0.019	&0.20	&5.41		&1		&3.14	&693.61	\\
\hline
\end{tabular}
\begin{tabbing}
Notes: 	\=[SJ] /[SJ]	\= \kill
Notes:	\>* 	\>Bubble beginning moved to a later time\\
		\>[JS]/[SJ]	\>From \cite{SJ:2001}/\cite{JS:2001}
\end{tabbing}
$t2c$ number of days from date of the fit until predicted crash date, $t2c = t_c$ - today\\
$\beta=0.01$ indicates that the optimal value of $\beta\le 0.01$\\
\textbf{Bold}  values of $\beta$ and $\omega$  are well outside the range specified in Eq.~\ref{lppl}\\
} 
\end{minipage}

\end{table}
\clearpage

\subsection{The `best' fits of the LPPL model }

We now fit the LPPL model to the raw data for each of the bubbles preceding the 11 crashes identified for the Hang Seng index over the period 1970 to 2008 (as selected by the criteria in Section~\ref{ss:what is a crash}). We use the minimum RMSE as the criterion for best fit. For each crash:

\LI 
\item The first line of Table~\ref{tab:Hang Seng_fits} shows the parameters of the LPPL model fit as given in the JS studies, but with the linear parameters $A, B$ and $C$ recalculated for time expressed in days rather than years.
As the RMSE was not reported for the JS studies (except for the LPPL fitted to the bubble preceding the 1997 crash) this too has been recalculated by us.

\item The second line shows the parameters for our best fit to the raw data.
The results are based on the raw data, despite our reservations about its appropriateness (Section~\ref{ss:index}), because we want to compare our results with those of the JS studies.\footnote{
For the crash of 1973 \cite{JS:2001} used the log instead of the raw index, so we report both log and raw fits specifically for that year.}

\item If this is not within the bounds for a crash prediction, then subsequent lines show the next best fit that is (or might be).
\li

Variation in the values of the critical parameters $\beta$ and $\omega$ sufficiently large to take them across their acceptable boundaries lead to only quite small fluctuations in the RMSE. This can bee seen, for example, for the crashes of 1973 and 1980 (see Table~\ref{tab:Hang Seng_fits}). 

We were interested in comparing our LPPL fits to those found in the JS studies. However, given the high sensitivity of the RMSE to small changes in the value of $\omega$ (see Section~\ref{ss:sensitivity}) and as the values for $\beta$ and $\omega$ were reported to only one decimal place in the JS studies, our re-calculated RMSEs will be different from those that were obtained in these studies.
We can see this in the bubble ending in the crash of 1997, where we have not only our recalculated RMSE using the parameters rounded to one decimal place, but also the RMSE using the unrounded parameter values as found by \cite{JLS:2000}; the latter fit is considerably better than our recalculation (RMSE=436 rather than 532 Hang Seng Index units). This improvement is almost certainly due to using the exact rather than the rounded value of $\omega$.
So caution needs to be taken when comparing the RMSEs for the fits reported in the JS studies and our fits.

Of the eight pre-crash bubbles fitted in the JS studies we find virtually the same parameters for the LPPL model for six of them; 
namely, those preceding the crashes of 1971, 1978, 1987, 1989, 1994 and 1997.
However, for their other two bubbles we found different parameters as follows:


\begin{itemize}

\item[1973:] For this bubble, \cite{SJ:2001} report the fit to the log of the Hang Seng index, rather than to the raw index. We have used both the log and the raw index. When we fit the log of the index we find a better fit than that reported in \cite{SJ:2001} with values of both $\beta$ and $\omega$ outside their acceptable ranges.
For comparison with other bubbles we also fitted the raw index;  we find that  the best fitting LPPL model has   a value for $\beta=0.26$, which is within the acceptable range of 0.15 -- 0.51, but for $\omega=1.45$, which is well below the lower bound of its critical range of 4.8 -- 8.0 (see Equation~\ref{lppl}).

\item[1980:] We were able to reproduce the fit reported in \cite{SJ:2001}, with a crash predicted 3 days later, but it was not the best fit that we found. 
	Our best fit predicted a crash after 52 days, and had critical parameter values $\omega= 7.51$, which is acceptable, but $\beta=0.01$, which is outside the acceptable range.

\end{itemize}

There are three pre-crash bubbles that were not considered in the JS studies; one, in 1981, they did not consider a crash (but see Section~\ref{ss:what is a crash}), and two others were later than their period:

\begin{itemize}

\item[1981:] We find a best fit for which both $\beta(=2.41)$ and $\omega(=3.02)$ are well outside their acceptable ranges.
	As $\beta > 1$,  this fit would have been rejected by the criteria used in the JS studies (see Section~\ref{s:fitting}).
	The first fit that has a $\beta <= 1$ has $\omega=4.75$, which is just acceptable, but with a $\beta=1$, i.e. no power law, so well outside its acceptable range.
	It might be argued that this peak was too soon (8 months) after the trough following the previous crash of 1980 for an LPPL model to be fitted on the grounds of there being insufficient data. But, as we have argued in Section~\ref{ss:what is a crash}, we believe it should have been.
	
\item[2000:] Our best fit to the bubble has both critical parameters $\beta (=1.00)$ and $\omega (=18.35)$ well outside their respective acceptable ranges.
	There is a fit that does have these parameters within their acceptable ranges, and predicts a crash after only 9 days; but it is not the best fit.

\item[2007:] Our best fit to this bubble has parameters well within the ranges required for a crash and the crash is predicted for the day it actually occurred.

\end{itemize}

\section{Conclusion}
\label{s:conclusion}
The LPPL model for pre-crash bubbles on stock markets, as reported in 
the JS studies, has important consequences. Our analysis has led us to the following conclusions.

The mechanism proposed to lead to the LPPL fluctuations as reported in  \cite{JLS:2000} must be incorrect as it requires the expected price to be increasing throughout the bubble (as  recognized later by \cite{SZ:2006}). In about half the studies they reported the LPPL model fitted to the index (or its log) decreases at some point during the bubble. Hence, either another explanation is required or the fits have to be redone with a constraint on the parameters that leads to LPPL fits that never decrease. 
Also, in the JS studies the fits were made to the raw rather than the log of the index for all but one (1973) of the eight bubbles, even though the assumption upon which the use of the raw rather than the log should be used was certainly {\bf not} met in six of these seven bubbles.
So, on both counts, these studies should no longer be used to support a conclusion that the proposed mechanism underlies the LPPL model.

Identifying crashes and bubble beginnings was not well specified in the JS studies. In particular, it is not clear why one peak, that of 1981, was not identified as a crash initiator. Moreover, moving the trough that marks the beginning of a bubble forward by `eye' in half the data sets is not really satisfactory. While we have taken more care in identifying those peaks that initiated crashes, we have still, for comparison, used the same bubble beginnings as used in the JS studies. In future, empirical studies need to establish a clear criterion for this procedure.

In the JS studies, the fits of the LPPL to the data were only accepted if the exponential parameter $\beta$ was $< 1$. That is, the fits showed an exponential {\em increase}. 
It would be stronger to reject the LPPL model if a $\beta \ge 1$ is found.

In our study the two critical parameters of the fitted LPPL models, $\beta$ and $\omega$, do fall within acceptable ranges in 7 of the 11 bubbles.
Of the remaining four bubbles, an LPPL model with critical parameters within their respective acceptable ranges could be found for all but one crash (1973).
However, these LPPL models did not have the best fits (minimum RMSE).
For one crash (1980) the best fit would be acceptable if the lower end of the acceptable range of $\beta$ was decreased, i.e. a range of 0.01 -- 0.51.
For another (1981), a fit with $\beta>1$ would also have to be ruled out to save the hypothesis. 
For two crashes (1973 and 2000), there seems to be no saving strategy. 
That the bubbles leading to the 1981 and 2000 crashes do not satisfy the criteria is particularly negative as these are two of the three crashes for which the ranges on the critical parameters were not set {\em post hoc} in the JS studies.

Finally, while the objection that with seven parameters a curve can be fitted to any data \cite{LPC:1999} is not directly relevant, since  no goodness of fit is measured here, it is indirectly highly relevant. The RMSE of the fit of the LPPL model (Eq. \ref{lppl}) to the data is highly sensitive to small but not to large fluctuations in one of the critical parameters ($\omega$); this makes the search for the LPPL that minimizes the RMSE unreliable. Moreover, substantial fluctuations in both parameters together can result in quite small changes in the RMSE \cite{BCP:2010}. 
This suggests that the permissible ranges for these parameters should not be independent of one another. 

Despite these criticisms, and because of the partial success of correctly predicting the 2007 crash, we believe that it is worth investigating whether fitted LPPL models with critical parameters in acceptable non-independent ranges can be used to give a probabilistic, rather than an all-or-none prediction of an impending crash. Furthermore, the use of the stock price alone is unlikely to be the only input for predicting stock market crashes. Using both trading volume and the log returns of stock prices in a spin model of heterogeneous agents, \cite{KBF:2008} are able to explain the origins of bubbles and crashes. Their approach, which appears promising, suggests a close correspondence between the magnetization of the spin model and trading volume, thereby enabling them to interpret the switch between bull and bear markets.

\newpage
\appendix
\begin{center}{\bf Appendices}\end{center}

\section{Derivation of Log Periodic Power Law}

To derive the LPPL from Eq. \ref{p=h}, substitute for $h$  as given in Eq. \ref{h2}: 

\begin{equation}
\nonumber
 \log p(t) = \kappa \int_{t_0}^t B'(t_c-t')^{-\alpha} \{1 + C' \cos(\omega \log(t_c-t') + \phi')\} dt'.
\end{equation}
 
\noindent Substituting $\beta$ for $1-\alpha$ and 
	$\psi(t')$ for  $\omega \log(t_c-t') + \phi' $ 
	and integrating gives:\footnote{Using Wolfram's Mathematica online integrator}

\begin{eqnarray}
\nonumber 
\log p(t) &=& - \kappa B' 
	\left[(t_c-t')^{\beta} 
		\left\{\frac{1}{\beta} +  \frac{C'}{\beta^2 + \omega^2 }
				 \left(\omega\sin \psi(t') +\beta \cos \psi(t') \right)
				 \right\}\right]^t_{t_0}	\\
\nonumber	
	&=& \kappa B' 
	\left[(t_c-t_0)^{\beta} 
		\left\{\frac{1}{\beta}+ \frac{C'}{\beta^2+\omega^2}(\omega \sin \psi(t_0) + \beta \cos \psi(t_0))
		\right\}
	\right.	\\
\nonumber 	&& \hspace*{1 cm}
	\left.	-(t_c-t)^{\beta}	
		\left\{\frac{1}{\beta}+ \frac{C'}{\beta^2+\omega^2}(\omega \sin \psi(t) + \beta \cos \psi(t))
		\right\}
	\right]. \\
\nonumber 
\therefore \log p(t_c) &=& \kappa B' (t_c-t_0)^{\beta} 
		\left\{\frac{1}{\beta}+  \frac{C'}{\beta^2+\omega^2}
			(\omega\sin \psi(t_0) + \beta \cos \psi(t_0) )
 		\right\} \\
\nonumber 
\therefore \log p(t) &=& 
		\log p(t_c) - \kappa B' (t_c-t)^{\beta}
			 \left\{\frac{1}{\beta} +\frac{C'}{\beta^2+\omega^2}
			 	\left(\omega\sin \psi(t) + \beta \cos \psi(t) \right)
			\right\} 	\\
\nonumber &=& \log p(t_c) - \frac{\kappa B'}{\beta} (t_c-t)^{\beta}
		\left\{1+\frac{\beta C'}{\sqrt{\beta^2+\omega^2}}\cos (\psi(t) + \phi'' )
		\right\}	\\	
\nonumber 	  
		&=&	A + B(t_c-t)^{\beta}\left\{1+C\cos(\omega \log(t_c-t) + \phi)\right\},
\end{eqnarray}
\noindent 
	where
	$A = \log p(t_c),\ 
	 B= -\kappa B' /\beta,\ $
	$C= \beta C'/\sqrt{\beta^2+\omega^2}$ and
	$\phi = \phi' +\phi''$,
which is the LPPL of Eq. \ref{lppl} with $y_t = \log(p_t)$.

\section{Search algorithm}

\fbox{\begin{minipage}{\textwidth}

\begin{enumerate}
\setcounter{enumi}{-1}
\item For each of the four parameters $\beta, \omega, t2c$ and $\phi$, fix the lower $L$ and upper $U$ bounds for the seeds.
	For a subset $\mathcal{P}$ of selected parameters ($\beta$ and $\omega$), fix the minimum width $W$ to continue searching.
\item Choose as the current seed $S1 \leftarrow (L + U)/2$, the mid point of the current lower and upper bounds.
\item	Run the unbounded Nelder-Mead Simplex search from the current seed $S1$, which will return a solution $S2$.
\item	Construct a hypercube in the space of $\mathcal{P}$ using $S1$ and $S2$, with their minimum as the bottom corner: $B \leftarrow \min(S1, S2)$; 
	and their maxima as the top corner: $T \leftarrow \max(S1, S2)$.
\item	For $p \leftarrow 1:\mbox{size}(\mathcal{P})$, i.e. for each of the selected parameters, do:
	\begin{itemize}
	\item[if] $B_p - L_p < W_p$ i.e. if there is too little space under the hypercube on the $p^{th}$ dimension in $\mathcal{P}$, set $B_p \leftarrow L_p$, i.e. set the bottom of the hypercube on the $p^{th}$ dimension to its lower bound,\\
		{\bf else} recursively search from step 1, with $L'\leftarrow L$ and $U' \leftarrow U, U'_p \leftarrow B_p$, i.e. search under the hypercube;
	\item[if] $U_p - T_p < W_p$, i.e. if there is too little space above the hypercube on the $p^{th}$ parameter, set $T_p \leftarrow U_p$, i.e. set the top of the hypercube on the $p^{th}$ parameter to its upper bound,\\
		{\bf else} recursively search from step 1, with $L' \leftarrow L, L'_p \leftarrow T_p$ and $U' \leftarrow U$, i.e. search above the hypercube.
	\end{itemize}
\end{enumerate}

\vspace{1 cm}
\begin{center}
	Initial bounds on the four parameters for selecting seeds
	
	\fbox{
		\begin{tabular}{llrrr}
						&$\beta$	&$\omega$	&$t2c$		&$\phi$	\\
						&		&rads 		&days		&rads\\
		\hline
		lower			&0		&0			&1			&0	\\
		upper			&2		&20			&260		&$\pi$	\\
		minimum width		&0.2		&2			&-			&-
		\end{tabular}
		}
\end{center}
\end{minipage}
}	

\clearpage 
\bibliographystyle{plain}

\bibliography{irfc}

\end{document}